\documentclass[twocolumn,amsmath,amssymb,aps, prm,nofootinbib]{revtex4-2}
\usepackage{graphicx}
\usepackage{color,soul}
\usepackage{amsmath,amssymb}
\usepackage{mathrsfs}
\usepackage{color}
\usepackage{afterpage}
\usepackage{pifont}
\usepackage[version=3]{mhchem}
\usepackage{natbib}
\usepackage{soul}
\usepackage[caption=false]{subfig}
\usepackage{array}
\usepackage{multirow}
\newcommand{\cmark}{\ding{51}}%
\newcommand{\xmark}{\ding{55}}%
\renewcommand{\thesection}{\Roman{section}} 
\usepackage[colorlinks, citecolor=blue,urlcolor=blue, linkcolor=blue, bookmarks=false]{hyperref}
\hypersetup{colorlinks=true , citecolor=blue, urlcolor=blue, linkcolor=blue}
\begin{document}

\title{Nonrelativistic spin splittings and altermagnetism in twisted bilayers of centrosymmetric antiferromagnets}

\author{Sajjan Sheoran} 
\email{phz198687@physics.iitd.ac.in}
\author{Saswata Bhattacharya}
\email{saswata@physics.iitd.ac.in}
\affiliation{Department of Physics, Indian Institute of Technology Delhi, New Delhi 110016, India}

\begin{abstract}
  Magnetism-driven nonrelativistic spin splittings (NRSS) are promising for highly efficient spintronics applications. Although 2D centrosymmetric (in four-dimensional spacetime) antiferromagnets are abundant, they have not received extensive research attention owing to symmetry-forbidden spin polarization and magnetization. Here, we demonstrate a paradigm to harness NRSS by twisting the bilayer of centrosymmetric antiferromagnets with commensurate twist angles. We observe $i$-wave altermagnetism and spin-momentum locking by first-principles simulations and symmetry analysis on prototypical MnPSe$_3$ and MnSe antiferromagnets. The strength of NRSS (up to 80 meV\AA) induced by twisting is comparable to SOC-induced linear Rashba-Dresselhaus effects. The results also demonstrate how applying biaxial strain and a vertical electric field tune the NRSS. The findings reveal the untapped potential of centrosymmetric antiferromagnets and thus expand the material's horizons in spintronics. 
\end{abstract}
%\pacs{}
%\keywords{DFT, valley drift, Berry curvature, circular dichroism}
\maketitle

Spin splittings in the electronic structure of crystalline solids play a pivotal role in spintronics applications (e.g., spin transistor)~\cite{vzutic2004spintronics,fert2008nobel}. The conventional spin-orbit coupling (SOC) induced Rashba-Dresselhaus ~\cite{dresselhaus1955spin,rashba1960properties,vas1979spin,bychkov1984properties} in nonmagnetic and Zeeman effects in ferromagnetic (FM) materials create spin splittings under certain crystalline (i.e., inversion [$\mathcal{P}$]) and time-reversal symmetry ($\mathcal{T}$) breaking~\cite{young2012spin}, respectively. SOC-induced spin splitting and resulting spin polarization engender spin-orbit torques~\cite{ciccarelli2016room}, while FM spin polarization has been widely known for spin generation and detection~\cite{vzutic2004spintronics}. However, the SOC effect introduces spin dephasing mechanisms~\cite{elliott1954theory,seitz1968solid,dyakonov1972spin}%, i.e., Elliott-Yafet~\cite{elliott1954theory,seitz1968solid} spin relaxation and Dyakonov-Perel~\cite{dyakonov1972spin} spin dephasing
, limiting the practical application. In addition, materials with heavy elements having significant SOC imparts additional challenges, including scarcity, toxicity, and instability. Therefore, nonrelativistic spin splitting (NRSS) is an important avenue to pursue.

Recently, antiferromagnetic (AFM) materials have emerged as viable substitutes for nonmagnetic and FM materials, benefiting from resilience towards stray fields, ultrafast dynamics, and magnetotransport effects~\cite{han2023coherent,baltz2018antiferromagnetic,jungwirth2016antiferromagnetic,jungwirth2018multiple,vzelezny2018spin}. The coupling of spin to lattice degrees of freedom via a staggered collinear compensated magnetism leads to alternating NRSS, termed altermagnetism~\cite{pekar1964combined, vsmejkal2022beyond, vsmejkal2022emerging}. Numerous efforts have been undertaken to investigate NRSS in AFM materials by breaking combined $\mathcal{P}\mathcal{T}\tau$ and/or $U\tau$ symmetries, where $U$ and $\tau$ are spinor and translation symmetry, respectively~\cite{hayami2019momentum,yuan2020giant,hayami2020bottom,yuan2021prediction,gonzalez2021efficient,vsmejkal2022giant,yuan2023degeneracy}. Nevertheless, the majority of AFM spin splittings are limited to bulk materials (e.g., MnF$_2$~\cite{dufek1993electronic,yuan2020giant}, LaMnO$_3$, and MnTiO$_3$~\cite{yuan2021prediction}), require SOC (e.g., MnS$_2$~\cite{yuan2021prediction} and ZnV$_2$O$_4$~\cite{maitra2007orbital}), or external perturbation~\cite{sivadas2016gate,sheoran2024multiple,zhao2022zeeman}.

Since the experimental revelation of 2D magnetic ordering, 2D vdW magnetic materials have garnered significant attention in scientific research, emerging as promising contenders for future information technology. Interestingly, two recent works focus on spin splitting in FM NiCl$_2$~\cite{he2023nonrelativistic} and FeBr$_2$ (although ``\textit{hidden}'')~\cite{yuan2023uncovering}  monolayers vdW stacked antiferromagnetically.  In contrast, antiferromagnetism-induced spin splitting among centrosymmetric materials with AFM order within each layer is not achieved due to $\mathcal{P}\mathcal{T}$ symmetry-enforced spin-degeneracy. Despite being abundant in nature, this impedes practical applications of 2D centrosymmetric AFM materials~\cite{basnet2022controlling,autieri2022limited, pournaghavi2021realization,aapro2021synthesis,sattar2022monolayer}.

This study generates NRSS and altermagnetism in $\mathcal{P}\mathcal{T}$-symmetric AFM monolayers vdW stacked with a relative twist. We perform density functional theory (DFT) simulations on twisted bilayer (tb-) MnPSe$_3$ and MnSe as prototypical candidates. The $i$-wave spin-momentum coupling arises in the 2D BZ for $\theta$ $(\neq 0^\circ, 60^\circ)$ tb-MnPSe$_3$ and MnSe. Based on the symmetry analysis, we find that the strengths of NRSS along specific crystallographic $k$-paths are comparable to the conventional SOC-induced Rashba-Dresselhaus effects. Moreover, external perturbations (i.e., electric and strain fields) provide exceptional tunability to NRSS.

\begin{figure}[h]
	\begin{center}
		\includegraphics[width=7.5cm]{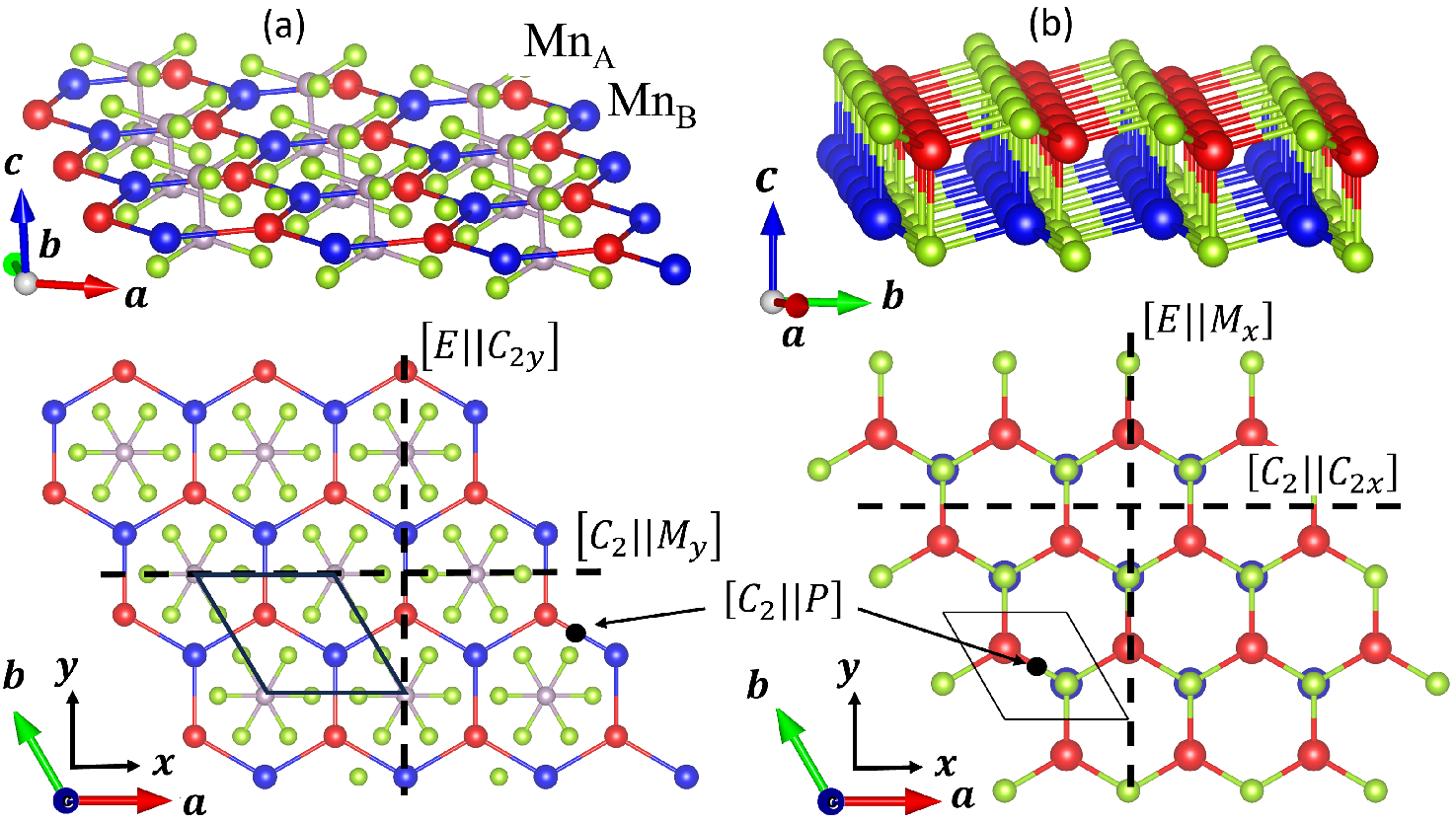}
	\end{center}
	\caption{Crystal structure of the monolayer (a) MnPSe$_3$ and (b) MnSe. The red and blue spheres indicate Mn atoms with the opposite collinear magnetic densities. The brown and green spheres represent P and Se atoms, respectively. The Cartesian ($x$, $y$, $z$) coordinate system and the hexagonal unit cell (with solid black lines) are shown for each case. The nontrivial spin-group symmetries are also highlighted. $E$ and $\mathcal{C}_2$ represent identity and two-fold rotation (about an axis perpendicular to spins) in spin-space, respectively. $\mathcal{M}_i$ and $\mathcal{C}_{2j}$ denote the mirror reflection perpendicular to the $i$ axis and the two-fold rotation parallel to the $j$ axis in real-space, respectively. $\mathcal{P}$ represents the real-space inversion. %Symmetry operations are calculated using using FINDSYM~\cite{stokes2005findsym}.
	}
	\label{str}
\end{figure}

MnPSe$_3$ and MnSe (space group \#162, $P\overline{3}1m$) represent two distinct classes of vdW materials that possess exceptional exfoliation properties~\cite{basnet2022controlling,autieri2022limited, pournaghavi2021realization,aapro2021synthesis,sattar2022monolayer}. Unlike the majority of other 2D magnetic materials, they exhibit an AFM arrangement, conforming to the conventional collinear N\'eel order on the honeycomb lattice [Fig.~\ref{str}]. This in-plane antiferromagnetism is different from the A-type antiferromagnetism observed in various other 2D vdW compounds, i.e., MnBi$_2$Te$_4$~\cite{zhao2021even}, CrI$_3$~\cite{sivadas2018stacking}, and CrSBr~\cite{lee2021magnetic}, where individual layers exhibit FM order but stack antiferromagnetically. The antiferromagnetism of MnPSe$_3$ is ``truly" in-plane and differs from that of MnSe, where Mn ions with opposite magnetic moments (Mn$_\textrm{A}$ and Mn$_\textrm{B}$) form unusual out-of-plane ordering within the individual layer. %The magnetic space group (MSG) of single-layer MnPSe$_3$ and MnSe is $C2'/m$ (type-III Shubnikov MSG), containing $\mathcal{P}\mathcal{T}$, $TC_{2i}$, and $M_{i}$ symmetry operations ($i$=$x$ for MnSe and $i$=$y$ for MnPSe$_3$ [see Fig~\ref{str}]).
Note that the orientation of on-site magnetic moments concerning the lattice only matters if SOC is included. Therefore, nonrelativistic spin-group formalism is described as the symmetry transformation of decoupled real and spin space~\cite{brinkman1966theory,litvin1974spin, litvin1977spin, vsmejkal2022beyond,liu2022spin}. The spin-symmetry operations $[R_i||R_j]$ of monolayer MnPSe$_3$ and MnSe are indicated in Fig.~\ref{str}, where the transformation on the left (right) of the double vertical bar acts on the only spin (real) space. In addition, collinear magnets always have additional symmetry $[\overline{\mathcal{C}}_2||T]$ arising from spin-only groups, where $\overline{\mathcal{C}}_2$ is the two-fold rotation perpendicular to the collinear spin axis, followed by spin-space inversion. Mn$_\textrm{A}$ and Mn$_\textrm{B}$ sublattices are connected through $[\mathcal{C}_2||\mathcal{P}]$ symmetry in monolayer MnPSe$_3$ and MnSe. $[\mathcal{C}_2||\mathcal{P}][\overline{\mathcal{C}}_2||T]$ ($\equiv$$\mathcal{P}\mathcal{T}$\footnote{Therefore, we use ``$[\mathcal{C}_2||\mathcal{P}]$'' and ``$\mathcal{P}\mathcal{T}$'' interchangeably.}) symmetry transforms energy eigenstate $E(k,\sigma)$ as $[\mathcal{C}_2||\mathcal{P}][\overline{\mathcal{C}}_2||T]E(k,\sigma)$=$[\mathcal{C}_2||\mathcal{P}]E(-k,\sigma)$=$E(k,-\sigma)$, leading to spin degeneracy throughout the Brillouin zone (BZ). We have verified that through DFT+U calculations performed on the projector augmented wave method~\cite{kresse1999ultrasoft} based VASP~\cite{kresse1996efficient} code (methods are detailed in Sect.~I of supplemental material (SM)~\cite{SuMa}). DFT simulated energy bands for  monolayer  MnPSe$_3$ and MnSe are doubly degenerate (see Sect.~II in SM~\cite{SuMa}). The semiconducting MnPSe$_3$ and MnSe have a magnetic moment of $\sim$4.5$\mu_B$/Mn with weak interlayer coupling. In addition, $[\mathcal{C}_2||\tau]$ can also enforce spin-degeneracy by connecting opposite spin sublattices by translation ($\tau$) as $[\mathcal{C}_2||\tau]E(k,\sigma)$=$E(k,-\sigma)$. Since 2D systems have only in-plane components of momentum $k_{||}$, nonrelativistic Hamiltonian for 2D systems may have symmetries other than $[\mathcal{C}_2||\mathcal{P}]$ and $[\mathcal{C}_2||\tau]$ enforcing spin degeneracy. For example, $[\mathcal{C}_2||\mathcal{M}_z]$ symmetry also enforces spin degeneracy throughout BZ in 2D materials, with $M_{z}:M_{z}k_{||}$=$k_{||}$ as a planer mirror reflection [see Sec. II of SM~\cite{SuMa} for details]. That makes achieving NRSS even more difficult for 2D materials. In the case of MnPSe$_3$ and MnSe monolayers, $[\mathcal{C}_2||\mathcal{M}_z]$ is already broken [Figs.~\ref{str}(a) and \ref{str}(b)], whereas type-III Shubnikov MSG ensures $[\mathcal{C}_2||\tau]$ symmetry-breaking%owing to the absence of screw axis, glide plane, and centering (body- or face-centering)
~\cite{dresselhaus2007group}. The only symmetry preserving spin degeneracy is $[\mathcal{C}_2||\mathcal{P}]$ symmetry in monolayers MnPSe$_3$ and MnSe.

\begin{figure}[t]
	\includegraphics[width=7.5 cm]{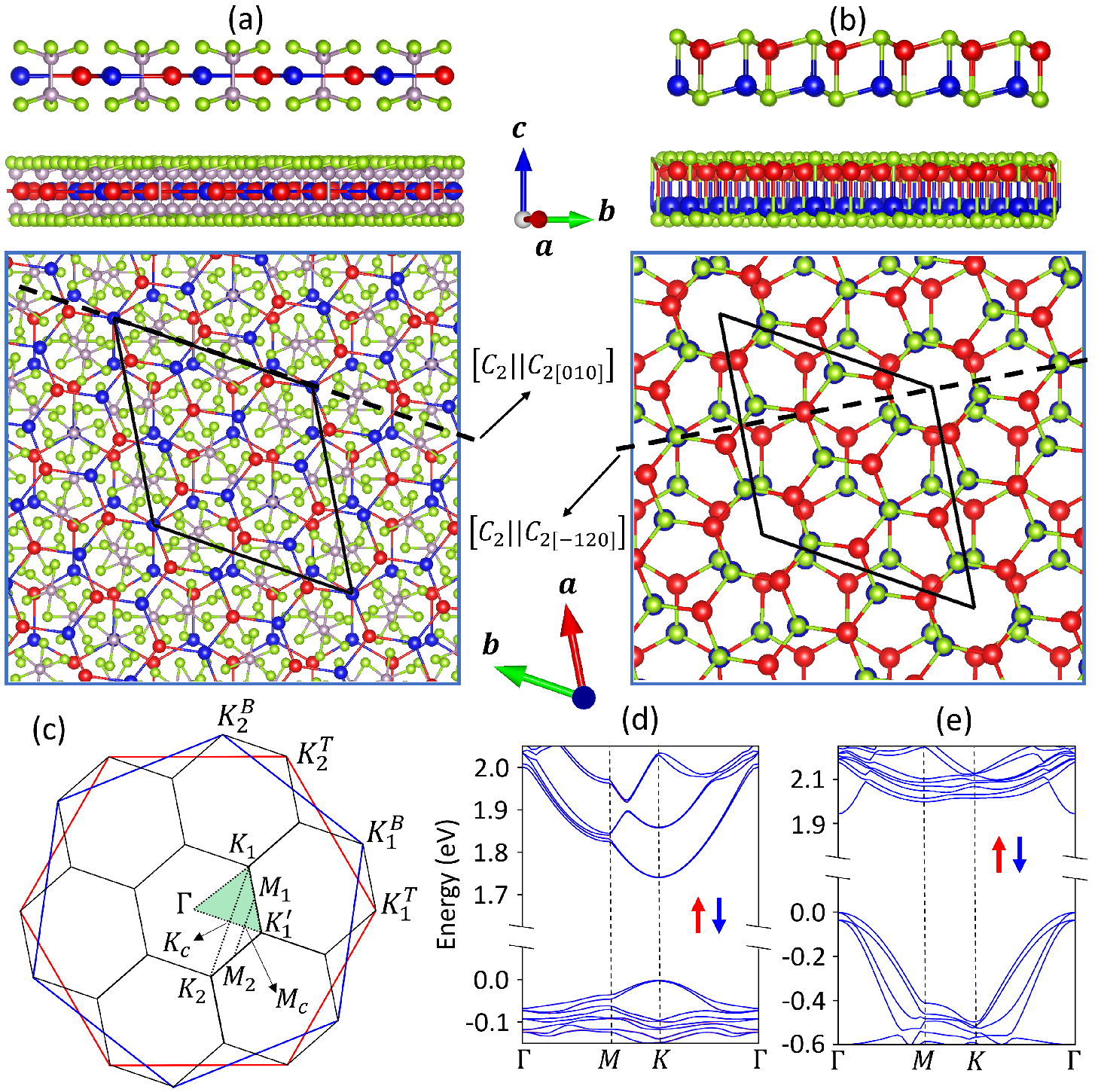}
	\caption{The Moir\'e superlattices formed by twisting bilayer of (a) MnPSe$_3$ and (b) MnSe by 21.79$^\circ$. (c) The Moir\'e BZ construction uses BZs of the top and bottom layers. The large red and blue hexagons are the first BZ of the top and bottom layers, respectively, and black hexagons represent the BZ corresponding to the Moir\'e superlattice. Spin-polarized band structure of (d) tb-MnPSe$_3$ and (e) tb-MnSe at the PBE level. The red and blue bands denote spin-up and spin-down states, respectively.}
	\label{pic1}
\end{figure}

Bilayer MnSe and MnPSe$_3$ are obtained from monolayers with various high-symmetry stackings as used in Ref~\cite{sheoran2024multiple}. Spin-up and spin-down states are degenerate for AA, AA$'$, AB, and BA stackings [see Sec. II of SM~\cite{SuMa}]. The $[\mathcal{C}_2||\mathcal{P}]$ symmetry enforces double degeneracy in AA, AB, and BA, whereas double degeneracy in AA$'$ stacking is protected by the $[\mathcal{C}_2||\mathcal{M}_{z}]$. Therefore, high-symmetry stackings are not an ideal for SOC-unrelated spin splitting in 2D $\mathcal{P}\mathcal{T}$-symmetric antiferromagnets.

Commensurate twisted bilayers are obtained using coincidence lattice theory~\cite{PhysRevB.90.155451} by taking the AA bilayer as the untwisted limit to break $\mathcal{P}\mathcal{T}$ symmetry. A periodic lattice structure, including Moir\'e superlattice, can form with special twist angle $\theta$, $\cos\theta=\frac{n^2+4mn+m^2}{2(m^2+mn+n^2)}$, where $m$, $n$ are whole numbers~\cite{dos2007graphene}. We only considered twist angles that resulted in reasonably sized commensurate supercells with the number of atoms in unit-cell fewer than 350. Figures~\ref{pic1}(a) and \ref{pic1}(b) show relaxed crystal structures and Moir\'e patterns in $\theta=21.79^\circ$ tb-MnPSe$_3$ and tb-MnSe [see Moir\'e BZ in Fig.~\ref{pic1}(c)]. Different possible interlayer and intralayer magnetic couplings $\uparrow \uparrow \uparrow \uparrow$, $\uparrow \downarrow \uparrow \downarrow$, $\uparrow \downarrow \downarrow \uparrow$, and $\uparrow \uparrow \downarrow \downarrow$ were considered to determine the preferred magnetic ordering [here, up and down arrows represent the relative magnetic moment direction on Mn atoms]. The most stable magnetic structure is $\uparrow \downarrow \uparrow \downarrow$, where magnetic order is intralayer and interlayer AFM [Figs.~\ref{pic1}(a) and \ref{pic1}(b)]. Twist angle leads to small variation in the magnitude of local magnetic moments from 4.441 to 4.448 $\mu_B$/Mn in tb-MnSe. The tb-MnPSe$_3$ and tb-MnSe are altermagnetic with opposite spin sublattices connected through the rotation symmetries ($[\mathcal{C}_2||\mathcal{C}_{2[010]}]$ and $[\mathcal{C}_2||\mathcal{C}_{2[-120]}]$, respectively) with net zero magnetization [Figs.~\ref{pic1}(a) and ~\ref{pic1}(b)]. In addition to Mn atoms, nonmagnetic ligands also contribute to $\mathcal{P}\mathcal{T}$ symmetry breaking in tb-MnPSe$_3$ and tb-MnSe. 

\begin{figure}[t]
	\includegraphics[width=8 cm]{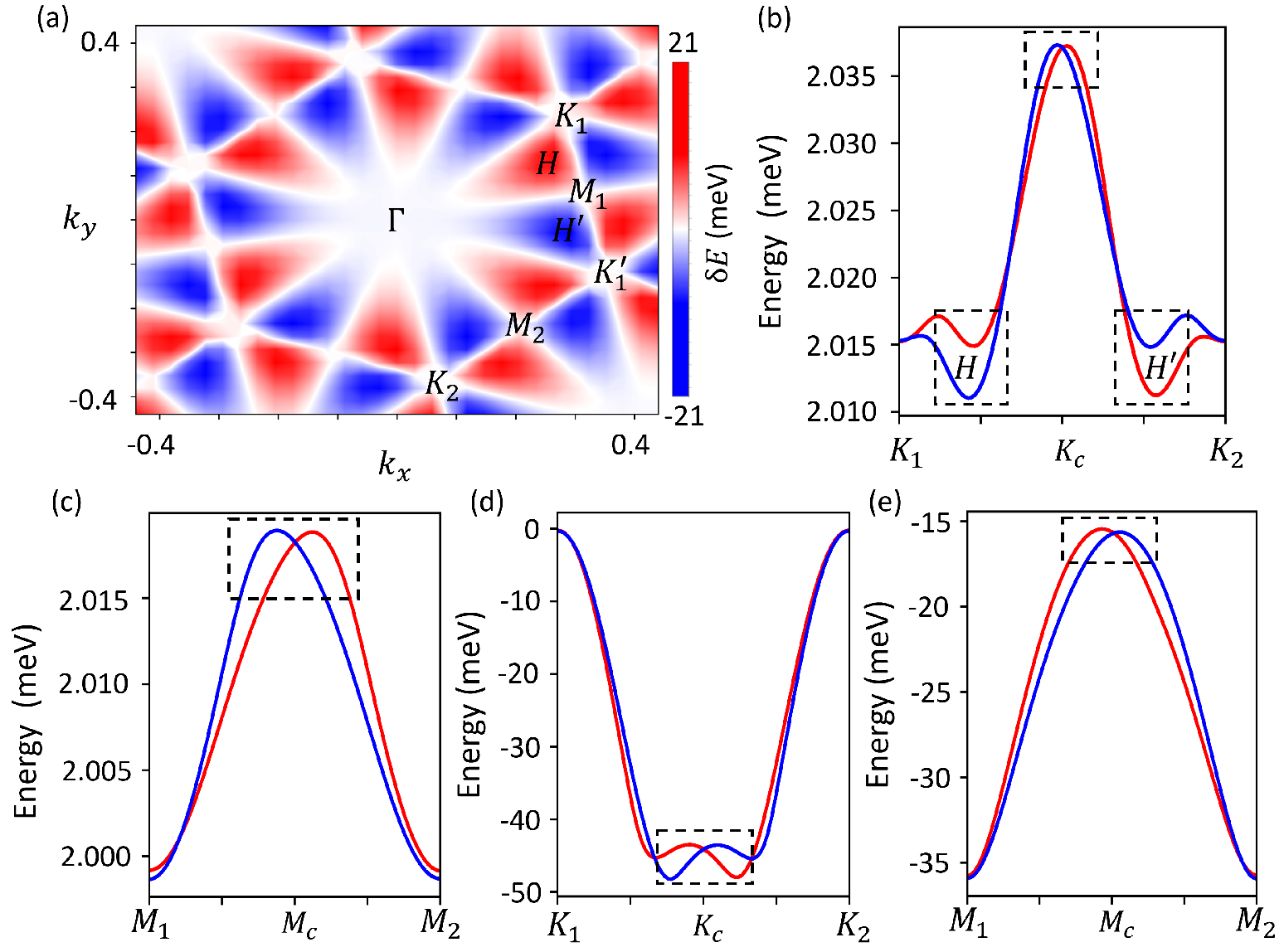}
	\caption{(a) Spin splitting energy [$\delta E=E_{\uparrow}(k)-E_{\downarrow}(k)$] distribution of valence band in 21.79$^\circ$ tb-MnSe. The units of $k_x$ and $k_y$ are \AA$^{-1}$. Conduction bands of tb-MnSe along the (b) $K_1$-$K_c$-$K_2$ and (c) $M_1$-$M_c$-$M_2$ paths [see Fig~\ref{pic1}(c) for paths]. Valence bands of tb-MnPSe$_3$ along the (d) $K_1$-$K_c$-$K_2$ and (e) $M_1$-$M_c$-$M_2$ paths. The red and blue curves denote spin-up and spin-down bands, respectively. Black dashed squares represent prominent spin splittings. Fermi energy is set to valence band maximum.}
	\label{pic2}
\end{figure}

Firstly, we compute the spin-polarized band structures of tb-MnPSe$_3$ and tb-MnSe along the high-symmetry paths (HSPs) [Figs.~\ref{pic1}(d),(e)]. The bands are spin degenerate along HSPs due to special symmetries arising at arbitrary $k$-point on HSP. For instance, $[\mathcal{C}_2||\mathcal{C}_{2[010]}]$ in tb-MnPSe$_3$ transforms spin-up to spin-down state along the $\Gamma$-$K$ path, enforcing degeneracy between them [see Sec. III of SM~\cite{SuMa}]. However, this is not the case for any generic $k$-point. No symmetry transform transform spin-up to spin-down at generic $k$-point, leading to the lifting of Kramers degeneracy. Therefore, the full BZ analysis of spin splitting is required. We plot spin splitting energy $\delta E$ [=$E_{\uparrow}(k)-E_{\downarrow}(k)$] of valence bands in tb-MnSe as a function of $k$ [Fig.~\ref{pic2}(a)]. The $\delta E$ is invariant under real-space inversion [$\delta E(k)=\delta E (-k)$] due to spin-only symmetry $[\overline{\mathcal{C}}_2||T]$, which transforms energy eigenstates $[\overline{\mathcal{C}}_2||T]E(k,\sigma)$=$E(-k,\sigma)$. Therefore, leading to 6-fold symmetric ($[E||\mathcal{C}_6]$) planar $i$-wave spin-momentum coupling, which is different from the 3-fold symmetry of SOC-induced $\delta E$ observed in well-known monolayer MoS$_2$~\cite{jafari2023robust}. %The higher-fold symmetry is attributed to the presence of additional symmetries (particularly $\mathcal{T}$) other than $3$-fold symmetry along $\Gamma$-$K$ and $\Gamma$-$M$ paths, which make $K_{1/2}$ ($M_{1/2}$) and $K'_{1/2}$ ($M'_{1/2}$) equivalent.
 Similar patterns are also observed for $\delta E$ of CB in tb-MnSe and VB in tb-MnPSe$_3$ (see Sect.~III in SM). The maximum NRSS is observed at the orthocenter ($H/H'$) of the triangle formed by $\Gamma$, $M$, and $K_1/K_1'$ points. Maximum splitting observed is 20.4, 4.2, and 5.1 meV for VB of tb-MnSe, CB of tb-MnSe, and VB of tb-MnPSe$_3$, respectively. Maximum $\delta E$ is smaller than well-known bulk antiferromagnets, i.e., MnF$_2$~\cite{yuan2020giant}, Fe$_2$TeO$_6$~\cite{zhao2022zeeman}, and LaMnO$_3$~\cite{yuan2021prediction}. The $\delta E$ observed in CB of tb-MnPSe$_3$ is negligible and beyond the accuracy of our calculations.

To understand the nature of NRSS, we plot band structures along both $K_1$-$K_c$-$K_2$ and $M_1$-$M_c$-$M_2$ directions [Figs.~\ref{pic2}(b)-(e)]. Interestingly, linear NRSS is observed around $K_c$ and $M_c$ for VB and CB of tb-MnSe and VB of tb-MnPSe$_3$. %Whereas splitting occurred around the $H/H'$ point predominantly of Zeeman-type in CB of tb-MnSe [Fig.~\ref{pic2}(b)]. 
The spin splittings exhibit contrasting characteristics at the $H$ and $H'$ points, featuring distinct valleys and maximum strength, suggesting the potential for valleytronics applications in twisted bilayers of antiferromagnets~\cite{sheoran2023probing,doi:10.1021/acs.jpcc.3c02819}. Note that the spin splittings around the $\Gamma$ point, along the $\Gamma$-$H/H'$ direction, exhibit cubic characteristics, which result in their being relatively small and, as such, are excluded from the current discussion~\cite{zhao2020purely}. 
\begin{figure}[b]
	\includegraphics[width=7 cm]{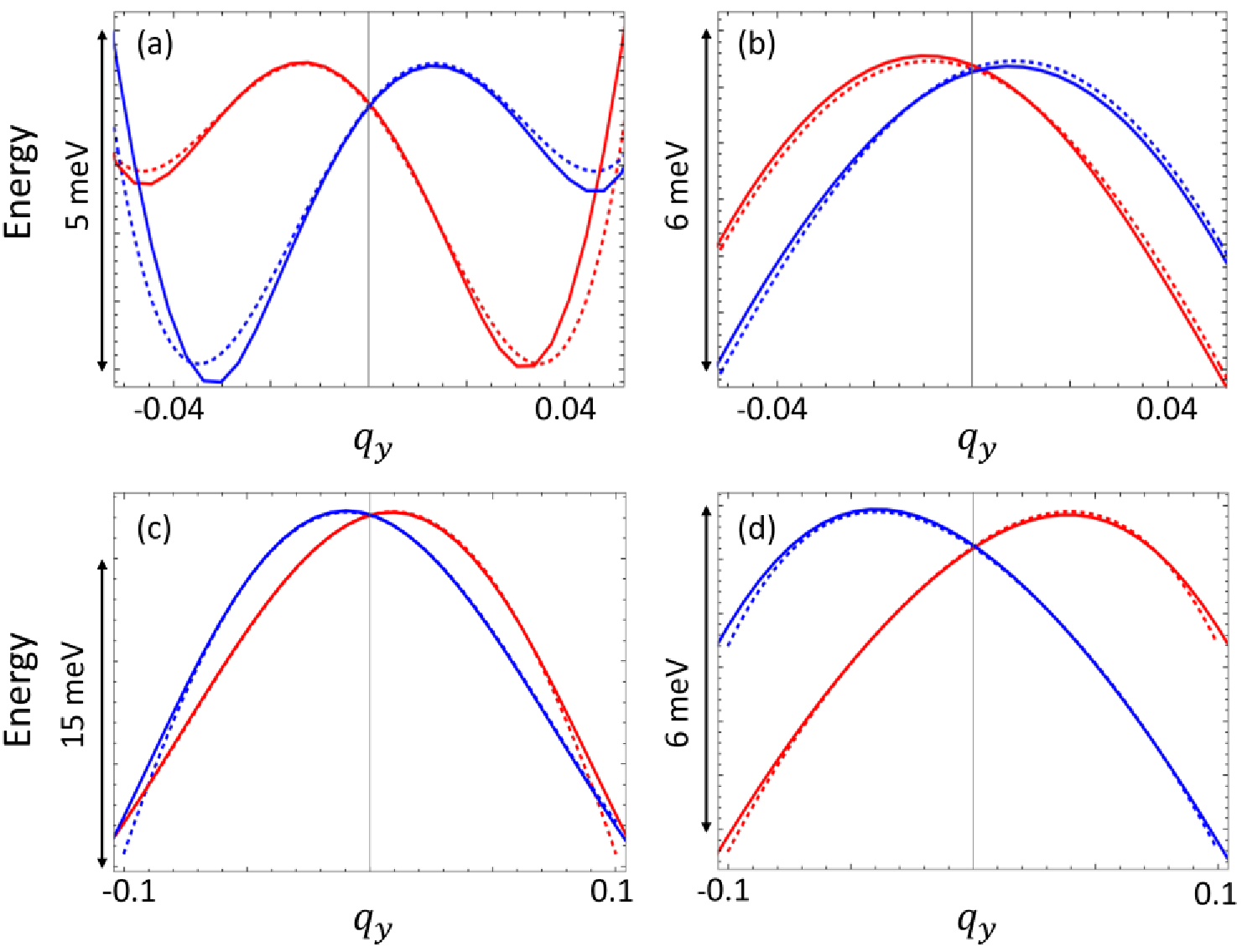}
	\caption{Band structures of tb-MnPSe$_3$ around (a) $K_c$ and (b) $M_c$ along $q_y$ direction. (c) and (d) are counterparts of (a) and (b), respectively, obtained for tb-MnSe. The solid and dotted lines are band structures obtained by DFT and the model described by Eq.~\ref{e1}, respectively.}
	\label{pic3}
\end{figure}
\begin{table*}[t]
	\caption{Classification of 2D materials based on the MSG type, magnetic order and their impact on the NRSS. The relevant spin-group symmetry is also indicated in the case of spin-degeneracy at generic $k$.}
	\centering
	\label{t1}
	\begin{tabular}{c|c| c| c| c |c}
		\hline
		\hline
		Spin splitting  &  \multicolumn{3}{c|}{Monolayer} & Twisted bilayer & Examples\\ \cline{2-5}
		prototype  & Magnetic order& MSG type & NRSS at generic $k$ &  NRSS at generic $k$ & \\ \hline
		 SST-1 & Nonmagnetic & II & \xmark($[\mathcal{C}_2||E]$) & \xmark($[\mathcal{C}_2||E]$) & MoS$_2$~\cite{naik2018ultraflatbands}, PtSe$_2$~\cite{xu2021evolution}\\
		SST-2 & Ferromagnetic &  I/III & \cmark & \cmark & NiCl$_2$, CrI$_3$, CrN, CrSBr~\cite{he2023nonrelativistic}\\
		SST-3 & Antiferromagnetic &  III & \xmark($[\mathcal{C}_2||\mathcal{P}]/[\mathcal{C}_2||\mathcal{M}_z]$) & \cmark & MnPSe$_3$, MnSe [\textcolor{blue}{This work}]\\
		SST-4 & Altermagnetic &  I/III  & \cmark  & - &\\
		SST-5 &  Antiferromagnetic & IV  & \xmark($[\mathcal{C}_2||\tau]$) & - &\\
		\hline
		\hline
	\end{tabular}
\end{table*} 

Spin splittings around $M_c$ and $K_c$ points are further analyzed using the symmetry-based model Hamiltonian, deduced using the ``\textit{method of invariants}''~\cite{tao2018persistent,zhao2020purely}. The symmetry element (besides identity) of $M_c$/$K_c$  point is $[\mathcal{C}_2||\mathcal{C}_{2[010]}]$ and $[\mathcal{C}_2||\mathcal{C}_{2[-120]}]$ for tb-MnPSe$_3$ and tb-MnSe, respectively [see Figs.~\ref{pic1}(a) and \ref{pic1}(b)]. The symmetry invariant terms include $\alpha q_{y'}\sigma_{z}$ and $q_{y'}q^2_{i}\sigma_{z}$ ($i=x', y'$), where $q=k-M_c/K_c$ are the momenta measured from $M_c$/$K_c$ [see Sec. III of SM~\cite{SuMa} for notation, derivation, and discussion]. Therefore, splitting is absent along the $q_{x'}$ ($K_c$-$K'_{1}$ and $M_c$-$K'_{1}$) direction, whereas it is present along the $q_{y'}$ direction ($K_c$-$K_{1/2}$ and $M_c$-$M_{1/2}$). To understand the NRSS along the $q_{y'}$ direction, it is possible to write an effective Hamiltonian ($H_{eff}$), up to third-order in $k$:
\begin{equation}
	H_{eff}=\alpha q_{y'}\sigma_{z}+\eta q_{y'}^3\sigma_{z}
	\label{e1}
\end{equation}
  $\alpha$ and $\eta$ are the constants determining the strength of NRSS. The primary linear term in Eq.~\ref{e1} leads to the linear splitting of spin-up and spin-down energy bands around $M_c$ and $K_c$ points, similar to the linearly split bands by SOC-induced Rashba and Dresselhaus effects. Note that spin splitting in Eq.~\ref{e1} originates from altermagnetic ordering and is completely nonrelativistic. On the other hand, the Rashba-Dresselhaus effect is induced by the spin-orbit field originating from noncentrosymmetric sites and is of relativistic origin. We fit the energy levels around $M_c$ and $K_c$ along the $q_{y'}$ direction to obtain spin-splitting parameters. The fits are obtained by minimization of the summation, $S=\sum_{i=1}^{2}\sum_{q}f(q)|\textrm{Det}[H_{eff}(q)-E^i(q)I]|^2$
over the $i^{th}$ energy eigenvalues [$E^i(q)$] as training sets. We have also included a weight function $f(q)$ with normal distribution to get a better fit near the spin-degenerate point and avoid overfitting. The obtained fits to the DFT energy levels of tb-MnPSe$_3$ and tb-MnSe are shown in Figs.~\ref{pic3}(a)-\ref{pic3}(d). The Hamiltonian in Eq~\ref{e1} with $\alpha$=58.6 meV{\AA} and $\eta$=34.2 eV\AA$^3$ provide the best fit to the VBs of tb-MnPSe$_3$ around the $K_c$ point [Fig.~\ref{pic3}(a)]. Whereas $\alpha$=39.8 meV{\AA} and $\eta$=3.5 eV\AA$^3$ are observed for VBs of tb-MnPSe$_3$ around the $M_c$ point, respectively [Fig.~\ref{pic3}(b)]. Similarly, linear splitting strength of 35.4 and 60.7 meV{\AA} is observed in CBs of tb-MnSe around the $K_c$ and $M_c$ points, respectively [Figs.~\ref{pic3}(c)-3(d)]. The NRSS is comparable to those experimentally reported in the literature (e.g., 10 meV{\AA} for KTaO3~\cite{varotto2022direct}, 4.3 meV{\AA} for LaAlO$_3$/SrTiO$_3$ interface~\cite{omar2022experimental}, $\sim$70 meV{\AA} in InGaAs/InAlAs interface~\cite{nitta1997gate}, and 77 meV{\AA} for MoSSe monolayer~\cite{hu2018intrinsic}). The growing field of twistronics makes NRSS observed in tb-MnPSe$_3$ and tb-MnSe experimentally accessible.

The 2D magnetic materials can be classified into five prototypes depending on the magnetic order, MSG, and whether NRSS is absent or present in a monolayer [Table~\ref{t1}]. Spin degeneracy in nonmagnetic materials (SST-1) is enforced by $[\mathcal{C}_2||E]$ and remains preserved under twisting operations. In contrast, FM materials (SST-2) show NRSS in both monolayer limits and two layers stacked antiferromagnetically with a twist~\cite{he2023nonrelativistic}. Altermagnetic materials (SST-4) have opposite-spin sublattices connected through mirror-rotation symmetries with opposite-spin electronic states separated in the momentum space. MSG type-IV always has AFM order with $[\mathcal{C}_2||\tau]$ symmetry (SST-5) and necessitates SOC to induce spin splitting~\cite{yuan2021prediction}. 2D AFM materials with MSG type-III containing  $[\mathcal{C}_2||\mathcal{P}]$ ($\mathcal{P}\mathcal{T}$) or  $[\mathcal{C}_2||\mathcal{M}_z]$ (SST-3) are unique, as NRSS is absent in the monolayer and presented in twisted bilayer. Therefore, the twisting operation generates splittings in SST-3 type materials, the most common magnetic ordering found in nature.
\begin{figure}[h]
	\includegraphics[width=7.5 cm]{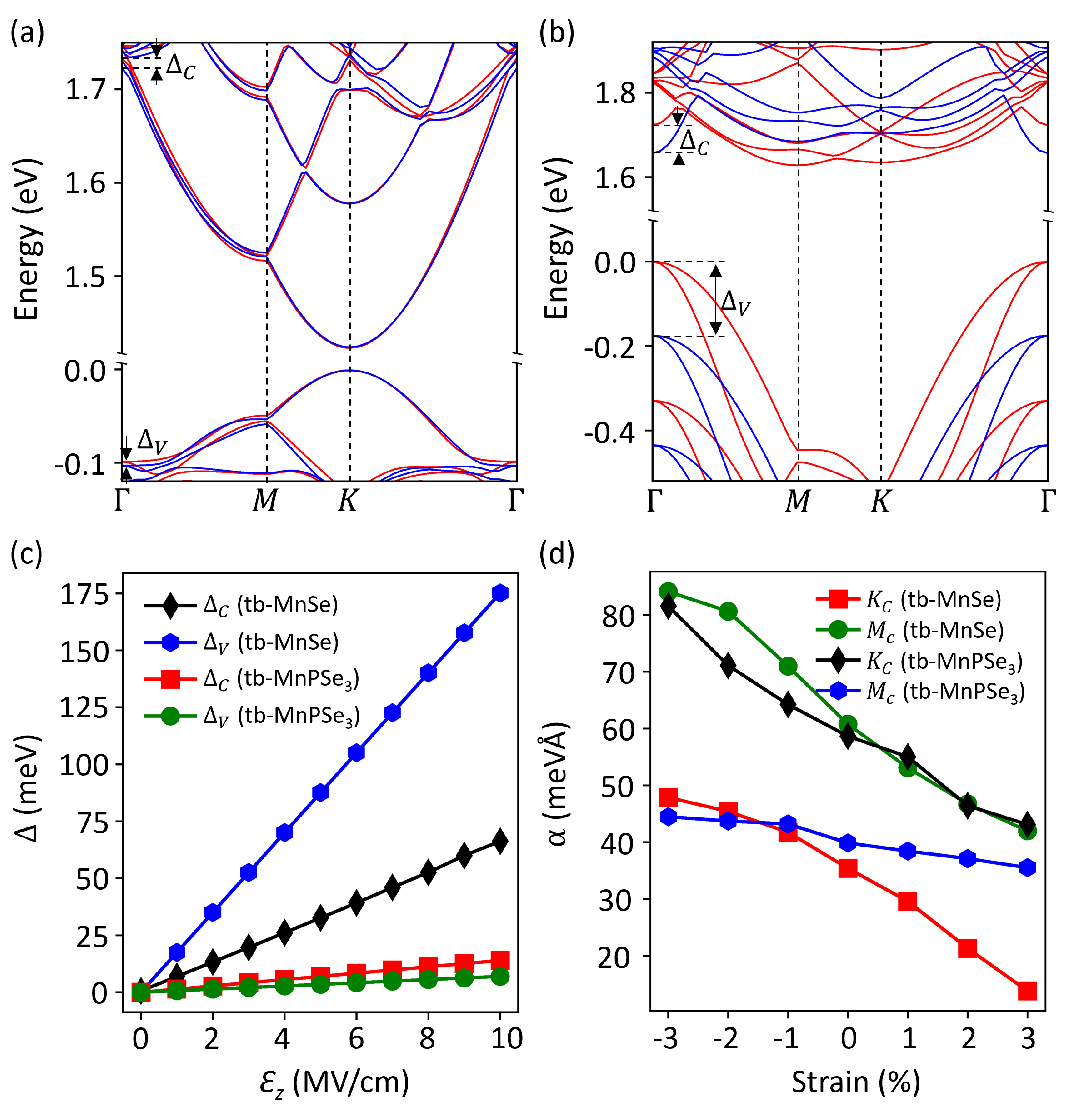}
	\caption{Band structures of 21.79$^\circ$ (a) tb-MnPSe$_3$ and (b) tb-MnSe in the presence of the out-of-plane electric field ($\mathcal{E}_z$) of strength 10 MV/cm. (c) The Zeeman spin splittings in the CB ($\Delta_C$) and VB ($\Delta_V$) of 21.79$^\circ$ tb-MnPSe$_3$ and tb-MnSe at $\Gamma$ point as a function of $\mathcal{E}_z$. (d) The variation in $\alpha$ (see Eq.~\ref{e1}) as a function of biaxial strain for 21.79$^\circ$ tb-MnPSe$_3$ and tb-MnSe.}
	\label{pic4}
\end{figure}

Although controlling crystal symmetries in bulk materials is challenging, it has been shown that gating can effectively break the symmetries in 2D materials, including twisted bilayers~\cite{weston2022interfacial,talkington2023electric,sheoran2023coupled}. In the following, we apply an out-of-plane electric field ($\mathcal{E}_z$) to the tb-MnPSe$_3$ and tb-MnSe in DFT simulations self-consistently using the approach introduced by Neugebauer and Scheffler~\cite{neugebauer1992adsorbate}. The electric field creates not only  polarization but also magnetization by breaking opposite spin-sublattice transformation through the magnetoelectric coupling~\cite{fiebig2005revival}. The Zeeman-like Hamiltonian under $\mathcal{E}_z$ are given by~\cite{zhao2022zeeman},
%\begin{equation}
%		\hat{H}_Z= \epsilon_{xy} \mathcal{E}_x \hat{\sigma}_{y} + \epsilon_{xz} \mathcal{E}_x \hat{\sigma}_{z}+ \epsilon_{yx} \mathcal{E}_y \hat{\sigma}_{x}+\epsilon_{zx} \mathcal{E}_z \hat{\sigma}_{x},
%		\label{e3}
%\end{equation}
$\hat{H}_Z= \lambda \mathcal{E}_z \sigma_{z}$,
where $\lambda$ is coefficient determining coupling strength. In the presence of $\mathcal{E}_Z$, the spin degenerate levels at the high symmetry points ($\Gamma$, $M$, and $K$) and along HSPs will be split into two sublevels, $E_+$=$\lambda \mathcal{E}_z$ and $E_-$=$-\lambda\mathcal{E}_z$ [Figs.~\ref{pic4}(a)-\ref{pic4}(c)]. We observe that the splitting induced by $\mathcal{E}_z$ in tb-MnPSe$_3$ and tb-MnSe exhibits markedly distinct characteristics. Specifically, an electric field $\mathcal{E}_z$ with a strength of 10 MV/cm results in nearly negligible splitting at the $\Gamma$ point for tb-MnPSe$3$, suggesting a small $\lambda$ [Figs.~\ref{pic4}(a),~\ref{pic4}(c)]. In contrast, for tb-MnSe, the $\Gamma$ point experiences a significantly larger Zeeman-type splitting ($\sim175$meV) induced by an electric field $\mathcal{E}_z$ of 10 MV/cm [Figs.~\ref{pic4}(b),~\ref{pic4}(c)]. This disparity can be explained through structural analysis: in tb-MnPSe$_3$, Mn atoms with opposite magnetic moments lie within the same $z$-plane, while in tb-MnSe, they are situated in different $z$-planes, thus supporting magnetoelectric coupling when an electric field is applied along the $z$-direction. On the contrary, when we compare the Zeeman splittings induced in the CB and VB of tb-MnSe at the $\Gamma$ point [Fig.~\ref{pic4}(b)], it becomes evident that the splitting in the VB is significantly greater in magnitude compared to that in the CB. This pronounced splitting in the VB of tb-MnSe can be attributed to the in-plane orbitals, which have wave functions segregated on different $z$-planes and, as a result, are more susceptible to the $\mathcal{E}_z$. In addition, tunability in electronic states can be achieved by the strain engineering of 2D materials~\cite{dai2019strain}. The in-plane biaxial strain preserves the crystal symmetry, thus creating no additional splittings. However, the strength of NRSS ($\alpha$) around $K_c/M_c$ for twisted bilayers are modified under biaxial in-plane strain [Fig.~\ref{pic4}(d)]. $\alpha$ increases with compressive strain and increases with tensile strain, providing exceptional tunability.

Similar effects were also investigated for other twist angles, including 9.43$^\circ$, 13.17$^\circ$, 27.79$^\circ$, 32.20$^\circ$, 38.21$^\circ$, and 42.10$^\circ$ (see Sect.~IV of SM~\cite{SuMa}). The $\delta E$ also depends upon the dispersiveness of energy bands, where $\delta E$ increases with increasing band dispersion. The linear NRSS is more prominent for the twist angles around $30^\circ$, as the structure deviates from the $\mathcal{P}\mathcal{T}$-symmetric ($\theta =0^\circ, 60^\circ$) counterparts by the highest amount. In addition, the strength of splitting is the same for twist angles $\theta$ and 60$^\circ$$-$$\theta$ (see Sect.~V of SM~\cite{SuMa}). MnPSe3 and MnSe contain relatively lighter elements with negligible SOC effects (see Sect.~VI of SM~\cite{SuMa}). The Zeeman splitting observed in bilayer MnSe with a twist angle of $\theta=0^\circ$ is $\sim$180 meV under 10 MV/cm of the vertical electric field~\cite{sheoran2024multiple}, nearly similar to 21.79$^\circ$ tb-MnSe of $\sim$175 meV with the same electric field. Similarly, the Zeeman effect in $0^\circ$ tb-MnPSe$_3$ is negligible~\cite{sivadas2016gate}, like 21.79$^\circ$ tb-MnPSe$_3$. Therefore, the order of Zeeman spin splitting depends much on how opposite spin-sublattices are arranged in the monolayer concerning the electric field and has less to do with the twist angle. Note that the models in this study include only spin degrees of freedom, thus revealing spin splitting qualitatively. For quantitative analysis, other degrees of freedom (i.e., orbital and sublattice) through first-principles or multiband tight-binding model calculations need to be included.

To summarize, we have shown that NRSS can be induced in 2D $\mathcal{P}\mathcal{T}$-symmetric antiferromagnets by taking bilayers with a relative twist. By first-principles calculations and symmetry analysis, we further predict  spin-moment coupling in 21.79$^\circ$ tb-MnPSe$_3$ and tb-MnSe that accommodate linear NRSS as large as $\sim$90 meV{\AA}. The lateral electric field split otherwise spin degenerate bands along the HSPs through magnetoelectric coupling, with more prominent effects in tb-MnSe. In addition, NRSSs are tunable using the biaxial strain. The measurement of these spin splittings can be conducted through well-established optical~\cite{sivadas2016gate} and electrical transport~\cite{shao2023neel} techniques commonly used in the field of spintronics. Employing antiferromagnets featuring spin-split bands as described in the present study may obviate the necessity for a heavy-metal layer, given that the current AFM mechanism yields a substantial magnitude of spin-moment splitting, even with lighter elements.  Moreover, the low-Z antiferromagnets with even larger NRSSs can be predicted by the inverse design approach with desired functionality~\cite{zunger2018inverse}. In addition, NRSS in Moir\'e-induced flat bands ($\theta$ $\lesssim$ $3^\circ$) can be an interesting avenue to prospect. We aspire to broaden the pool of available materials and enrich the field of AFM semiconductor spintronics~\cite{vzutic2004spintronics,fert2008nobel} through the complete realization of original devices.

\textit{Acknowledgments}.\textemdash S.S. acknowledges CSIR, India, for the senior research fellowship [grant no. 09/086(1432)/2019-EMR-I]. S. B. acknowledges financial support from SERB under a core research grant (grant no. CRG/2019/000647) to set up High-Performance Computing (HPC) facility ``Veena" at IIT Delhi for computational resources.

\bibliography{ref}

\end{document}